\documentclass[twocolumn,floatfix,superscriptaddress,aps]{revtex4}
\usepackage[dvips]{graphicx}
\DeclareGraphicsExtensions{.eps}
\begin{document}
\title{Selecting a single orientation for millimeter sized graphene sheets}

\author{R. \surname{van Gastel}}
\affiliation{University of Twente, MESA$^+$ Institute for Nanotechnology,
P.O. Box 217, NL-7500AE Enschede, The Netherlands}

\author{A.T. \surname{N'Diaye}}
\affiliation{II. Physikalisches Institut, Universit\"{a}t zu K\"{o}ln,
Z\"{u}lpicher Stra{\ss}e 77, 50937 K\"{o}ln, Germany}

\author{D. \surname{Wall}}
\affiliation{Department of Physics and Center for Nanointegration
Duisburg-Essen (CeNIDE), Universit\"{a}t Duisburg-Essen, D-47057 Duisburg,
Germany}

\author{J. \surname{Coraux}}
\affiliation{Institut N\'{e}el/CNRS, B\^{a}t. D, 25 Rue des Martyrs,
F-38042 Grenoble Cedex 9, France}

\author{C. \surname{Busse}}
\affiliation{II. Physikalisches Institut, Universit\"{a}t zu K\"{o}ln,
Z\"{u}lpicher Stra{\ss}e 77, 50937 K\"{o}ln, Germany}

\author{F.-J. \surname{Meyer zu Heringdorf}}
\affiliation{Department of Physics and Center for Nanointegration
Duisburg-Essen (CeNIDE), Universit\"{a}t Duisburg-Essen, D-47057 Duisburg,
Germany}

\author{N. \surname{Buckanie}}
\affiliation{Department of Physics and Center for Nanointegration
Duisburg-Essen (CeNIDE), Universit\"{a}t Duisburg-Essen, D-47057 Duisburg,
Germany}

\author{M. \surname{Horn von Hoegen}}
\affiliation{Department of Physics and Center for Nanointegration
Duisburg-Essen (CeNIDE), Universit\"{a}t Duisburg-Essen, D-47057 Duisburg,
Germany}

\author{T. \surname{Michely}}
\affiliation{II. Physikalisches Institut, Universit\"{a}t zu K\"{o}ln,
Z\"{u}lpicher Stra{\ss}e 77, 50937 K\"{o}ln, Germany}

\author{B. \surname{Poelsema}}
\affiliation{University of Twente, MESA$^+$ Institute for Nanotechnology,
P.O. Box 217, NL-7500AE Enschede, The Netherlands}

\date{\today}

\begin{abstract}
We have used Low Energy Electron Microscopy (LEEM) and Photo Emission Electron
Microscopy (PEEM) to study and improve the quality of graphene films grown
on Ir(111) using chemical vapor deposition (CVD). CVD at elevated temperature
already yields graphene sheets that are uniform and of monatomic thickness.
Besides domains that are aligned with respect to the substrate, other 
rotational variants grow. Cyclic growth exploiting the faster growth and
etch rates of the rotational variants, yields films that are
99 \% composed of aligned domains. Precovering the substrate
with a high density of graphene nuclei prior to CVD yields
pure films of aligned domains extending over millimeters. Such films can 
be used to prepare cluster-graphene hybrid materials for catalysis or
nanomagnetism and can potentially be combined with lift-off techniques
to yield high-quality, graphene based electronic devices.
\end{abstract}

\maketitle

Graphene potentially constitutes a new material for electronic
circuitry with vastly improved transport properties over
traditional silicon \cite{novoselov}. A large scale application of graphene
crucially hinges on a fabrication method that yields perfect graphene
sheets, that is low-cost and reliable. CVD growth of graphene on metals 
has recently been demonstrated to yield large graphene sheets of
uniform monatomic thickness \cite{alpha1,johann,sutter,loginova}.
This form of epitaxial CVD, which occurs through ethylene decomposition
on the uncovered parts of the metal substrate, is self-limiting since the
nature of the process limits the thickness of the graphene sheets to a
single atomic layer, in contrast to e.g. graphene formed on heated SiC
substrates \cite{tromp,emtsev}. Metal CVD thus appears to be the route
of choice for fabrication of large graphene sheets. In situ growth studies
with LEEM and PEEM have however, highlighted a new problem. The orientation
of the domains that make up the graphene sheet is not always in registry
with the substrate. For the case of e.g. Ir(111), four different
orientations have been observed \cite{loginova2}. The electronic properties
of graphene sheets depend sensitively on the relative orientation with respect
to the substrate \cite{zagreb,kellyprl,kellyprb}. Also, applications of
cluster superlattices of magnetically or catalytically active materials
grown on the graphene sheets \cite{alpha2} require full control over the
orientation of the domains. Here, we address this problem by
tailoring the epitaxial process to grow a millimeter sized, monatomic
thickness graphene sheet of single, aligned orientation.

An Ir(111) single crystal was heated to 1123 K and exposed to
a $1\cdot 10^{-7}$ mbar partial pressure of O$_2$ to remove
residual carbon contamination. CVD grwoth of graphene sheets was
performed by exposing the surface to ethylene at elevated temperatures.
Growth of the graphene ceases when the fractional surface coverage
of graphene approaches 1 ML. Threshold PEEM images using a Hg
discharge lamp yield a high intensity from the graphene flakes and
very low intensity from the bare Ir(111) surface. Contrast between
the different rotational domains is achieved in LEEM mode at various electron
energies.

\begin{figure}
\includegraphics*[width=8.5cm]{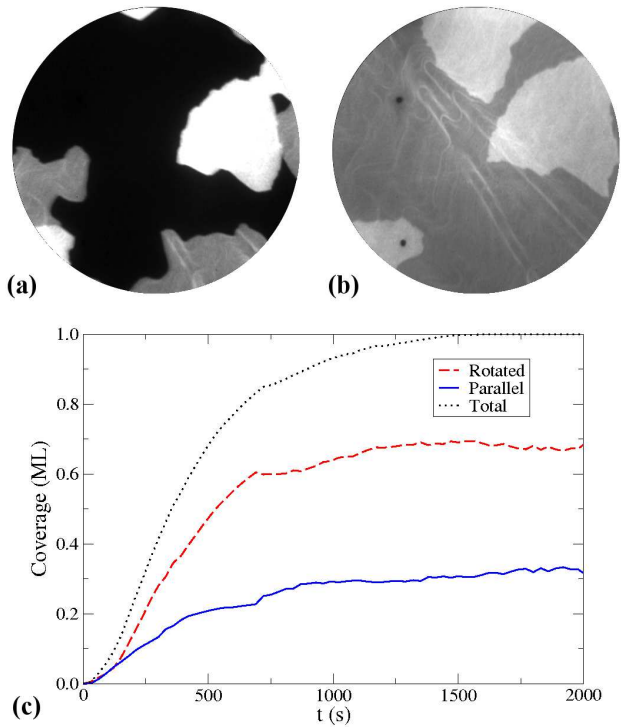}
\caption{50 $\mu$m Field Of View (FOV) PEEM images of graphene growth on
Ir(111) at T = 1411 K. {\bf (a, {\it t} = 265 s)} Two different types of
domains are observed to form. The brighter of the two is aligned with the
substrate, whereas the other, darker, type of domain is rotated by
approximately 30$^\circ$ with respect to the substrate.
{\bf (b, {\it t} = 2120 s)} The film has fully closed to form a graphene
sheet consisting of various rotational domains. {\bf (c)} Relative area
fractions of the two different types of domains that are visible in panels
(a)-(b). The majority of the graphene sheet consists of rotated domains.}
\label{fig1}
\end{figure}

First, growth of a graphene sheet was studied at a temperature of
1411 K by exposing to an ethylene partial pressure of
$1\cdot 10^{-7}$ mbar. The formation of the graphene is shown in
Fig. \ref{fig1}.
Initially, only a single phase forms that has its lattice vectors parallel
to the substrate lattice. Later, graphene domains that are rotated with
respect to the substrate lattice are observed to form at the edges of the
original nuclei and grow at a rate that is substantially faster. In what
follows, we shall refer to these as aligned and rotated domains,
respectively. The structure of the graphene sheet after it has completed
is shown in Fig. \ref{fig1}(b). From the simple observation that contrast
between different domains is observed in these  threshold PEEM images, we
conclude that there is a significant variation of the electronic
structure of the film between different domains. Even though the sheet
thickness is very uniform and could already be characterized as a high
quality graphene film, further control over the rotational orientation
of the domains is desired.

\begin{figure}
\includegraphics*[width=8.5cm]{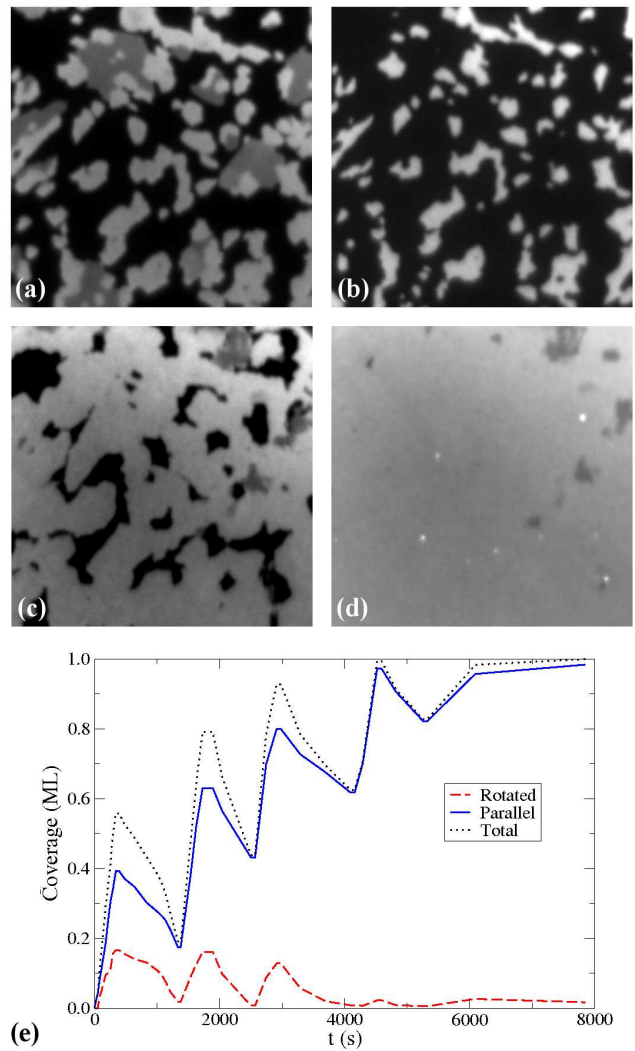}
\caption{102 $\mu$m FOV PEEM images of graphene growth on Ir(111) obtained
at a temperature of 1126 K. {\bf (a, {\it t} = 344 s)} Several parallel
and rotated domains have nucleated after the first growth cycle.
{\bf (b, {\it t} = 265 s)} The first O$_2$ etching step has completely
removed all rotated domains. {\bf (c, {\it t} = 795 s)} After two more
etch and growth steps, the graphene film now consists of a majority of
parallel domains. Rotated domains continue to nucleate at every growth
step, as is visible in the image. {\bf (d, {\it t} = 2120 s)} The film
has fully closed to form a graphene sheet. A relative fraction of
99 \% of the sheet consists of parallel domains. {\bf (e)} Relative
area fractions of the two different types of domains that are visible
in panels (a)-(d).}
\label{fig2}
\end{figure}

One way to produce a high quality graphene film of a single rotational
phase is done by exploiting the higher reactivity of the edges of the
rotated domains. Not only do the three types of rotated domains grow
at a rate that is higher than that of aligned domains, they are also
etched away by oxygen at an increased rate \cite{fjmzh}.
Fig. \ref{fig2} highlights this experimental approach.
The Ir(111) surface was alternatingly exposed to ethylene and O$_2$ at
partial pressures of $5\cdot 10^{-8}$ mbar. Exposure to ethylene
leads to the formation of new nuclei and continued growth of
aligned domains. It also gives rapid growth of any rotated
domains that have formed. Exposure to O$_2$, shown in Fig. \ref{fig2}(b),
then preferentially etches away the rotated domains until only parallel
domains remain. Cyclic repetition of this procedure yields a monolayer
thick, near uniform graphene sheet. A relative fraction of 99 \% of the
sheet consists of aligned domains. The drawback of this method is that
closure of the film to produce a ``perfect'' graphene sheet is not
possible. The last step in the growth process always has to be a growth
step, implying that the nucleation of rotated domains can not be fully
prevented.

\begin{figure}
\includegraphics*[width=8.5cm]{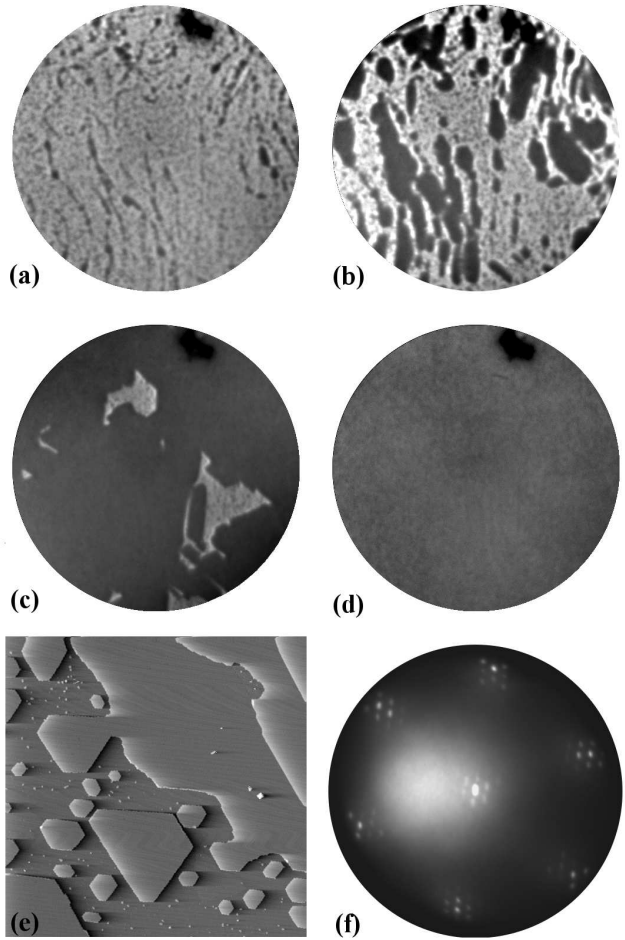}
\caption{4 $\mu$m FOV LEEM images of graphene growth on Ir(111) at a
temperature of 1113 K and recorded with an electron energy of 18.6 V.
The dark spot in the top of the images is an MCP defect.
{\bf (a, {\it t} = 0 s)} Start of ethylene exposure of the surface. The Ir(111)
surface has been precovered with many small graphene nuclei, appearing dark
in the LEEM images. {\bf (b, {\it t} = 139 s)} Several of the
predeposited domains have started to grow. Those domains that have edges along
the substrate crystallographic directions are not observed to grow.
{\bf (c, {\it t} = 631 s)} The graphene film has nearly evolved into a
sheet. No rotated domains that would yield a higher intensity than the
aligned domains, are observed. Several precovered domains still persist
and do not grow. {\bf (d, {\it t} = 1279 s)} The film has fully closed to form
a perfectly aligned graphene sheet.
{\bf (e)} A 0.25 $\mu$m FOV STM image of the growth of the parallel phase
taken after the graphene film was only partially completed. The smaller nuclei
with straight edges running along substrate crystallographic directions have
not grown significantly, whereas domains with edges of different orientation
have (I$_t$ = 0.5 nA, V$_t$ = 0.5 V). {\bf (f)} $\mu$LEED pattern obtained of
the closed graphene film. The orientation of the graphene is unaltered when
the beam is scanned over an area of several millimeters.}
\label{fig3}
\end{figure}

The nucleation of rotated domains occurs at the edges of parallel
domains \cite{loginova2}. In our measurements, we also observe that
growth of aligned domains and the nucleation of rotated domains occurs
predominantly at those edges that do not run parallel to the substrate
lattice vectors. This observation was exploited to further improve the
quality of the films beyond what was demonstrated in Fig. \ref{fig2}
with the cyclic recipe. A monolayer of ethylene was preadsorbed on the
surface at room temperature. Upon heating the substrate to the growth
temperature this leads to the formation of a high density of small
aligned graphene domains that have edges parallel to the substrate
lattice \cite{johann}. This effectively forces any graphene domains
that impinge on existing nuclei to maintain their aligned orientation.
Fig. \ref{fig3} highlights the subsequent growth when the substrate
is exposed to an ethylene partial pressure of $1\cdot 10^{-7}$ mbar.
Figs. \ref{fig3}(b) and (c) illustrate that growth of aligned domains is
observed only in those locations where domain edges are rough, having
an orientation deviating from the dense packed substrate directions.
Small domains with edges oriented parallel to substrate lattice vectors
are not observed to grow, illustrated by the STM image shown in
Fig. \ref{fig3}(e). Rotated domains do not form. The $\mu$LEED pattern
that is shown in Fig. \ref{fig3}(f) is measured over several millimeters
of our 6 mm wide sample. Defects are sporadically found, but always in
locations where we have to presume that they were induced by features
present on the Ir(111) substrate. The graphene sheet that is formed
through this recipe has the added advantage that its orientation is
uniquely determined by the orientation of the Ir(111) substrate. 

In conclusion, we have grown millimeter sized, graphene films of a
single orientation. Cyclic growth of the graphene film exploiting
the different growth and O$_2$ etching speeds of the domain variants
yields films that are aligned to the substrate dense packed orientation
up to a fraction of 99 \%. The final approach, using preadsorption
of ethylene on the Ir(111) surface at room temperature, followed by
CVD growth at elevated temperatures yields perfectly aligned sheets
that are ready for application.

\begin{acknowledgments}
Financial support through Deutsche Forschungsgemeinschaft is
gratefully acknowledged.
\end{acknowledgments}

\end{document}